%% file: main.tex
\documentclass{PoS}

\title{LHCb: Status and Prospects on the $b$ Anomalies}

\ShortTitle{$b$ Flavour Anomalies}

\author{\speaker{A.\ Hicheur}\thanks{also at U.~Constantine, Algeria.}\\
        LHCb Collaboration / Federal University of Rio de Janeiro.\\
        E-mail: \email{Adlene.Hicheur@cern.ch}}

\abstract{Since the start of the Large Hadron Collider program, direct searches for Beyond Standard Model (BSM) particles have constrained their mass scale to limits which are now above the energy reach of the current collider. As a result, studies of indirect probes of BSM physics have gained a considerable momentum, both experimentally and theoretically. The flavour anomalies in $b$ hadron decays are now recognized as an important laboratory for the indirect detection of BSM physics. This short review presents several key analyses in this area, and some prospects with future data.}

\FullConference{%
  BEAUTY2020\\
  21-24 September 2020\\
  Kashiwa, Japan (online)}

\usepackage{ifthen} % for conditional statements
\newboolean{pdflatex}
\setboolean{pdflatex}{true} % False for eps figures 

\newboolean{articletitles}
\setboolean{articletitles}{true} % False removes titles in references

\newboolean{uprightparticles}
\setboolean{uprightparticles}{false} %True for upright particle symbols

\newboolean{inbibliography}
\setboolean{inbibliography}{false} %True once you enter the bibliography

\input{preamble} % Add in the predefined LHCb symbols

\begin{document}
%\linenumbers
\section{Introduction}
Heavy Flavour decays are usually described by low energy effective Hamiltonians forming an Effective Field Theory (EFT) (see, e.g. reference \cite{Buchalla} for a review). The Hamiltonians are written as:
\begin{equation}
H = \sum_{i} V_{CKM}^i C_i(\mu) O_i(\mu),
\end{equation}
where $C_i(\mu)$ are the Wilson coefficients integrating out the physics above the scale $\mu$ (short range), $O_i(\mu)$ are current operators which matrix elements represent the low energy (non-perturbative/long range) hadronic physics, and $\mu$ is the renormalization scale (typically $\sim 1 \gev$) distinguishing the two regimes. $V_{CKM}^i$ represents the flavour coupling associated to an operator $O_i$, i.e., Cabibbo-Kobayashi-Maskawa (CKM) matrix elements for SM operators. The Wilson coefficients thus represent the quantities which are impacted by the intervention of BSM physics.\\
For the semileptonic tree decays, the coupling of the mediating $W$ boson does not discriminate between lepton flavours in the SM. On the contrary, a BSM mediator might exhibit different couplings between light and heavy leptons. This is referred to as Lepton Flavour Universality Violation (LFUV). Such an effect could also occur for the semileptonic loop decays $b\to s\ell\ell$ where the dominant operators are $O_7,~O_9,~O_{10}$. Loop decays could also be the ground of new dynamics involving Lepton Flavour Violation (LFV) where leptons of different flavours are produced together. A review of various LHCb analyses is proposed. Several of them exhibit deviations from SM that can be explained consistently with theoretical models.
\section{Semileptonic tree decays}
\begin{figure}[htb]%bt\centering
\begin{center}
\includegraphics[width=0.4\textwidth]{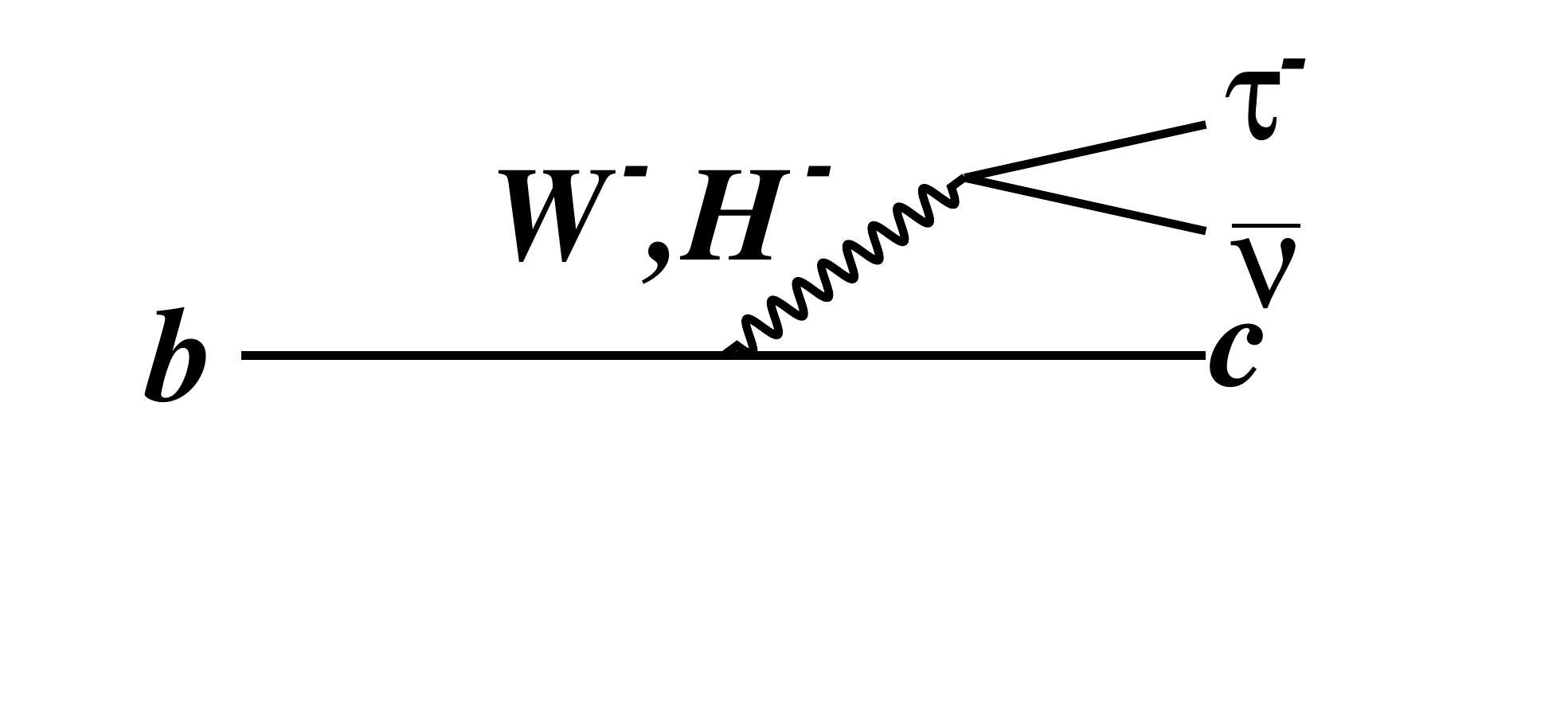}
\end{center}
\vspace{-1.3cm}
\caption{$b\to c\tau^-\bar\nu$ semileptonic transition with a mediating $W$ or charged Higgs boson as derived from 2HDM models discussed e.g. in Ref.\cite{Higgsrev_2HDM}.}
\label{Fig:btoctau_Higgs}
\end{figure}

The dominant decays ($b\to c \ell^- \bar\nu$) of this kind are generically written as $H_b\to H_c \ell^- \bar\nu$ where $H_b$ is a $b$ hadron and $H_c$ is a charm hadron. The search for a possible LFUV is performed through the measurement of the ratio:
\begin{equation}
R(H_c) = \frac{\mathcal{B}(H_b\to H_c \tau^- \bar{\nu})}{\mathcal{B}(H_b\to H_c \mu^- \bar{\nu})},
\end{equation}
where $\mathcal{B}$ denotes the branching fraction. A BSM mediating heavy boson might couple preferentially to the tau lepton, as in Fig.\ref{Fig:btoctau_Higgs}, and thus produce a $R(H_c)$ ratio different from the expected SM-based calculations.

\begin{figure}[htb]%bt\centering
\begin{center}
\includegraphics[width=0.59\textwidth]{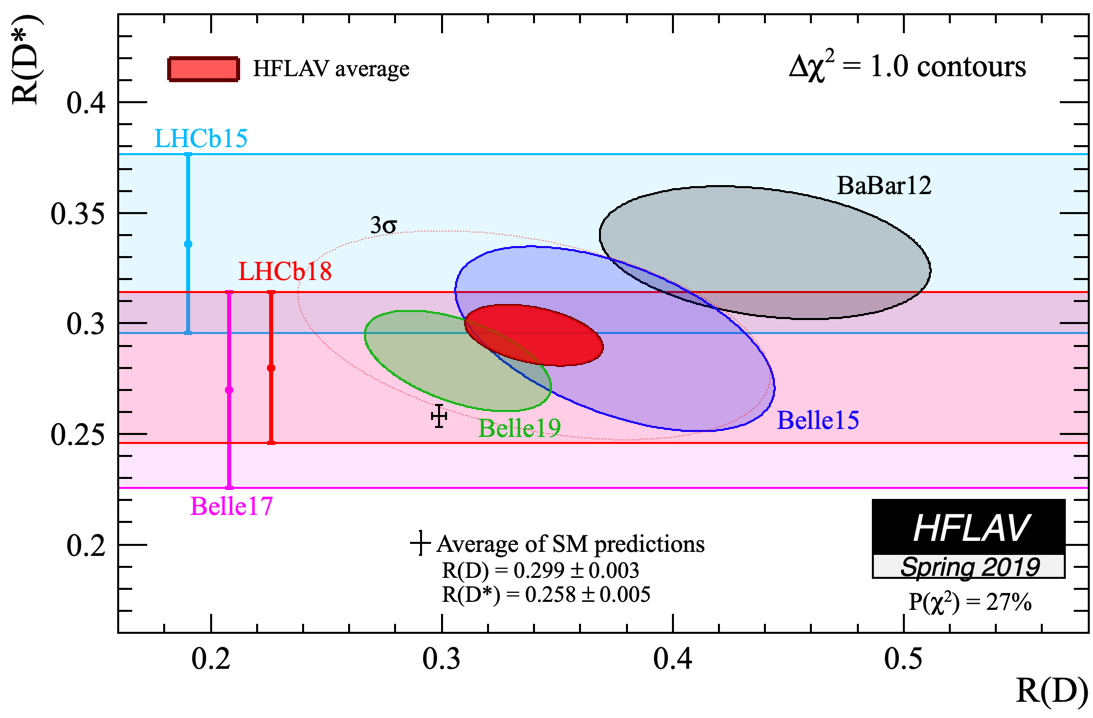}
\end{center}
\vspace{-0.7cm}
\caption{Average of $R(D)$ and $R(D^*)$ \cite{hflav}.}
\label{Fig:rd_rdstar_hflav}
\end{figure}

The modes $\bar{B}^0 \to D^{+(*)} \ell^- \bar\nu$ have drawn attention both at the $b$ factories and the LHCb experiment. The LHCb collaboration has been focusing so far on $R(D^{*+})$, where $D^{*+}$ is reconstructed via $D^{*+}\to D^0(\to \Km\pip)\pip$. The $\tau$ lepton is reconstructed in the muonic mode, $\tau^-\to\mu^-\bar{\nu}_\mu\nu_\tau$ \cite{Dstar_MuMode}, or the hadronic mode $\tau\to\pi\pi\pi(\pi^0)\nu_\tau$ \cite{Dstar_HadMode}. The discriminating variables include the missing mass, $m_{miss}^2 = (P_B-P_{D^*}-P_\mu)^2$, the momentum transfer $q^2 = (P_B-P_{D^*})^2$, the muon energy $E_\mu^*$, the $\tau$ lifetime (for the hadronic mode) and a Boost Decision Trees classifier \cite{BDT} to reject double-charm decays of the type $B \to D D X$ (for the hadronic mode). 
%To recover the $B$ four-momentum, the neutrino momentum needs to be determined. As LHCb is not a $4\pi$ detector, only a geometric method can be used, as depicted in Fig.\ref{Fig:pneut}. The component $p_\parallel(\nu)$ is inferred from the second order equation derived from the condition $P_B^2 = m_B^2$. The ambiguity in the solution is solved by a regression method \cite{RegressionMethod}. 
The muonic tau analysis \cite{Dstar_MuMode} obtained a measurement of $R(D^{*+})=0.336\pm0.027(stat)\pm0.030(syst)$ while the hadronic tau study \cite{Dstar_HadMode} gives $R(D^{*+})=0.291\pm0.021(stat)\pm0.026(syst)\pm0.013(BF)$, where the last uncertainty is due to the external branching fraction of the normalizing channel $\bar{B}^0\to D^{*+}\pim\pip\pim$. The latest HFLAV averaging \cite{hflav} in the $R(D)-R(D^*)$ plane, including the recent Belle collaboration $R(D^{(*)})$ measurements \cite{belle_rd_2019}, is shown in Fig.\ref{Fig:rd_rdstar_hflav}. Compared to an averaged series of SM-based predictions \cite{rd_theory}, a discrepancy of 3.1$\sigma$ is observed.
%\begin{figure}[htb]%bt\centering
%\begin{center}
%\includegraphics[width=0.4\textwidth]{pneut.pdf}
%\end{center}
%\vspace{-0.5cm}
%\caption{Sketch of the flight vector from the primary vertex (PV) to the $B$ vertex (BV) and the momenta of the visible daughters $\mu$ and $D$. The momentum component perpendicular to the $B$ flight vector, $p_\perp$, is used to infer the neutrino momentum.}
%\label{Fig:pneut}
%\end{figure}

A similar measurement with the decays $B_c^- \to \jpsi \ell^- \bar\nu$, $R(\jpsi)$, has been performed recently by the LHCb collaboration for the $\tau$ muonic mode \cite{RJpsi}, leading to \mbox{$R(\jpsi)=0.71\pm0.17(stat)\pm0.18(syst)$} which lies 2$\sigma$ above the range of the known theoretical estimates \cite{RJpsi_theory}.

\section{$b\to s\ell\ell$ transitions}
\begin{figure}[htb]%bt\centering
\begin{center}
\includegraphics[width=0.3\textwidth]{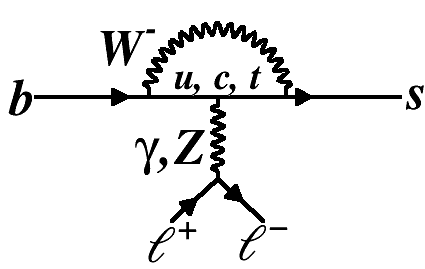}
\includegraphics[width=0.4\textwidth]{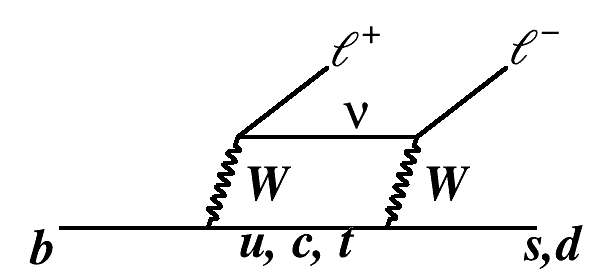}
\end{center}
\vspace{-0.5cm}
\caption{$b\to s\ell\ell$ (left) penguin and (right) box transitions.}
\label{Fig:btosll_diags}
\end{figure}

At quark level, these transitions proceed through the diagrams shown in Fig.\ref{Fig:btosll_diags}. The operators contributing to these decays are not evenly distributed in the $q^2 = m_{\ell\ell}^2$ range: at low $q^2$, $O_7$ dominates (for transitions to non-scalar hadrons), in the central $q^2$ region below the charmonium resonances, $O_7$ and $O_9$ interfere, and at high $q^2$ $O_9$ and $O_{10}$ interfere.
%\cite{lhcb_kmumu,lhcb_phimumu,lhcb_kstmumu,lhcb_lzmumu}
At the hadron level, the modes investigated by LHCb are $\Bp\to K^{*+}\ell^+\ell^-$, $B^0\to K^{0}\ell^+\ell^-$, $B^0\to K^{*0}\ell^+\ell^-$, $\Bs\to \phi\ell^+\ell^-$, and $\Lb\to \Lz\ell^+\ell^-$. 

\begin{figure}[htb]%bt\centering
\begin{center}
\includegraphics[width=6cm, height=4cm]{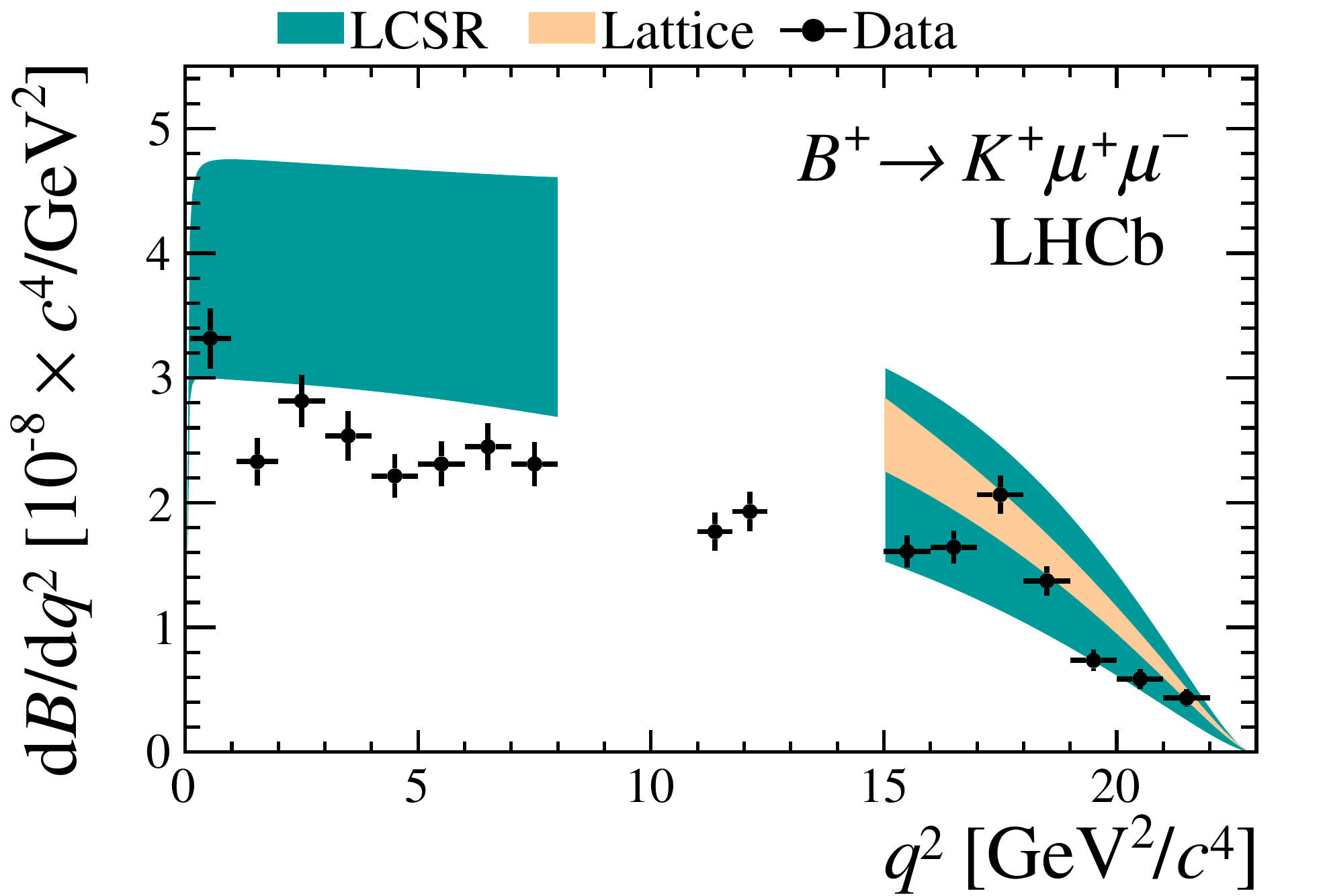}
\includegraphics[width=6cm, height=4cm]{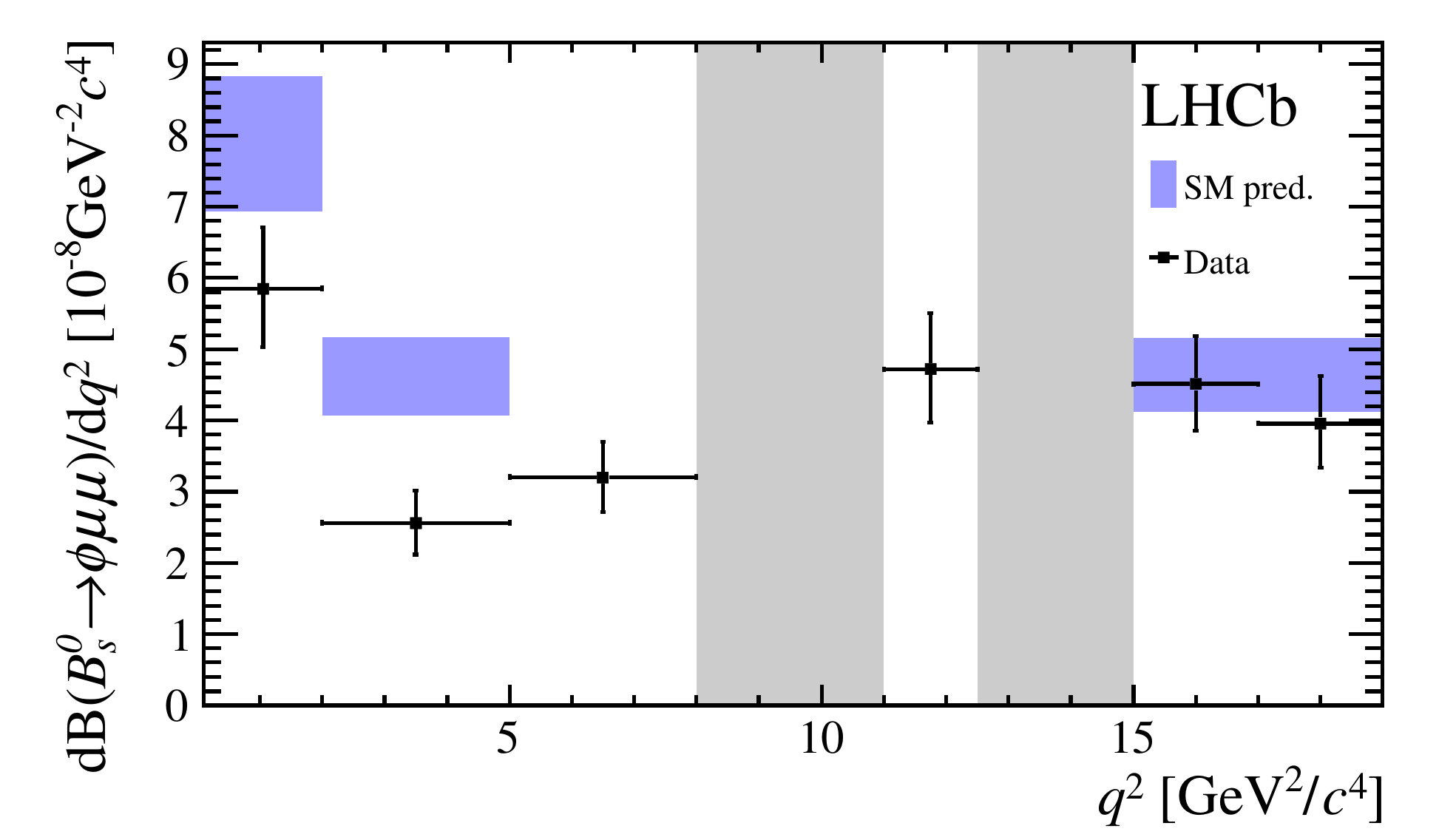}
\end{center}
\vspace{-0.5cm}
\caption{$\frac{d\Gamma}{dq^2}$ distribution for (left) $B^+\to K^{+}\mup\mun$ and (right) $\Bs\to \phi\mup\mun$. The \jpsi and \psitwos $q^2$ regions are vetoed. The points represent the data measurements and the rectangles or band represent the SM-based predictions.}
%\caption{$\frac{d\Gamma}{dq^2}$ distributions for (top left) $\Bp\to\Kp\mup\mun$, (top right) $B^0\to K^{0}\mup\mun$, (bottom left) $B^0\to K^{*0}\mup\mun$ and (bottom right) $\Bs\to \phi\mup\mun$. The \jpsi and \psitwos bands are vetoed.}
\label{Fig:dGdq2}
\end{figure}

\begin{figure}[htb]%bt\centering
\begin{center}
\includegraphics[width=6cm, height=4cm]{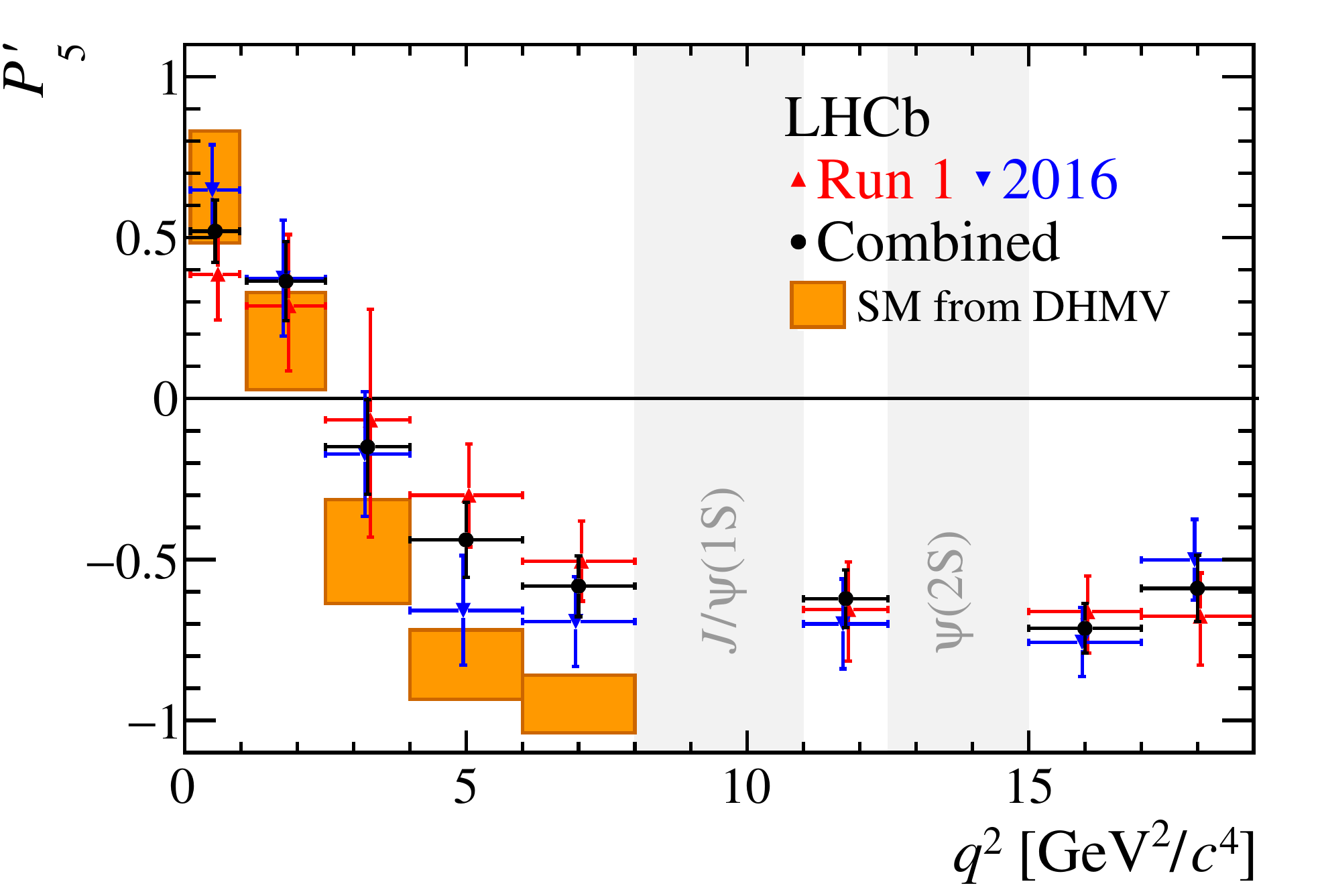}
\includegraphics[width=0.4\textwidth]{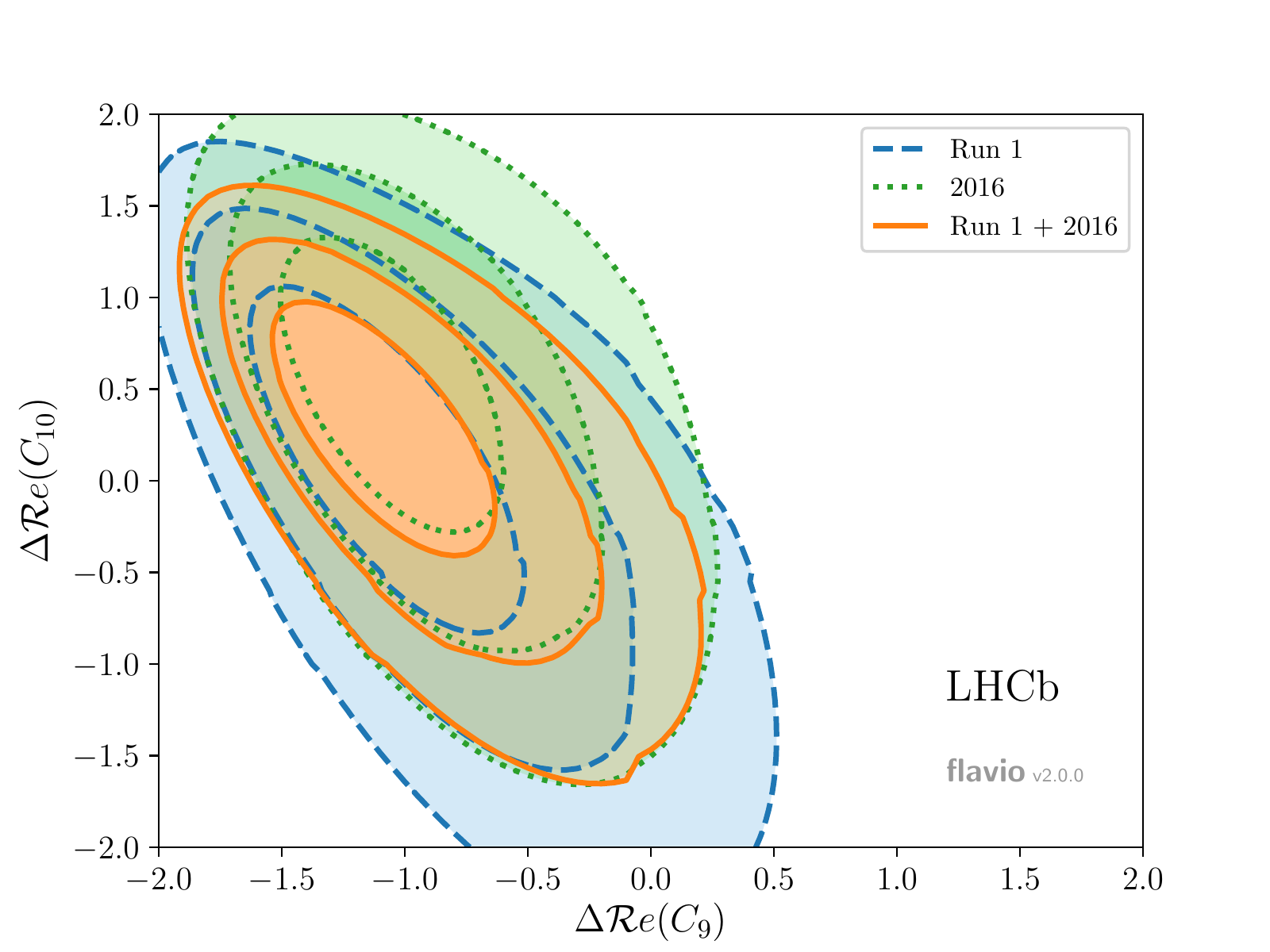}
\end{center}
\vspace{-0.5cm}
\caption{(left) Evolution of $P_5^\prime$ (see text) as a function of $q^2$ for $B^0\to K^{*0}\mup\mun$ and (right) resulting 1,2,3 $\sigma$ contours, using all the angular variables, of the deviations from SM of the real parts of the $C_9$ and $C_{10}$ Wilson coefficients.}
\label{Fig:P5Prime}
\end{figure}

A series of studies \cite{lhcb_xmumu} have dealt with the dynamics of the muonic modes, $\ell=\mu$, to infer the differential decay rate $\frac{d\Gamma}{dq^2}$, as illustrated in Fig.\ref{Fig:dGdq2}. The data is systematically below the SM-based theoretical predictions, with local discrepancies exceeding 3$\sigma$. Attempting to explore this intriguing behaviour, angular analyses were performed for $B^0\to K^{*0}\mup\mun$ \cite{lhcb_xmumu}(c), $\Bs\to \phi\mup\mun$ \cite{lhcb_xmumu}(b) and $\Lb\to \Lz\mup\mun$ \cite{lhcb_lzmumu_ang}. Quantities such as $P_5^\prime = \frac{S_5}{\sqrt{F_L(1-F_L)}}$ have been built to reduce the hadronic uncertainties \cite{descotes} from the coefficients $S_5$ and $F_L$ (fraction of the $K^*$ longitudinal polarization) of the angular distribution. For $B^0\to K^{*0}\mup\mun$, the discrepancy reported in previous studies for $P_5^\prime$ seems to be persistent as shown in Fig.\ref{Fig:P5Prime}(left). The fit for the deviations from SM to the real parts of the $C_9$ and $C_{10}$ Wilson coefficients gives the results depicted in Fig.\ref{Fig:P5Prime}(right). Considering only $C_9$, the deviation from SM is determined to be 3.3$\sigma$.

Another way to probe the presence of New Physics is to measure the ratio $R_X = \frac{\mathcal{B}(H_b\to X \mu^+\mu^-)}{\mathcal{B}(H_b\to X e^+e^-)}$, where $X$ denotes a hadronic system comprising a strange quark. The LHCb collaboration studied $R_K$ ($\Bp\to\Kp\ell^+\ell^-$) \cite{lhcb_rkplus_ratio}, $R_{K^*}$ ($B^0\to K^{*0}\ell^+\ell^-$) \cite{lhcb_rkstar_ratio} and $R_{pK}$ with the decay $\Lb\to p\Km\ell^+\ell^-$ \cite{lhcb_rpk_ratio}. For $R_K$, the explored $q^2$ range is $[1.1,6]~\gevgevcccc$, i.e. below the charmonium radiative tails, and above backgrounds of the type $\Bp\to\Kp\phi(\to\ell^+\ell^-)$. The $R_{K^*}$ analysis uses two bins in $q^2$, $[0.045,1.1]~\gevgevcccc$ (above the photon pole) and $[1.1,6]~\gevgevcccc$. Finally, $R_{pK}$ is measured with the requirements $q^2\in[0.1,6.0]~\gevgevcccc$ and $m(pK)<2.6\gevcc$.
%A notable fact in these analyses is that compared to the muonic decays, the electronic modes suffer from Bremsstrahlung tails as illustrated in Fig.\ref{Fig:Kll_masses}.

\begin{figure}[htb]%bt\centering
\begin{center}
\includegraphics[width=0.4\textwidth]{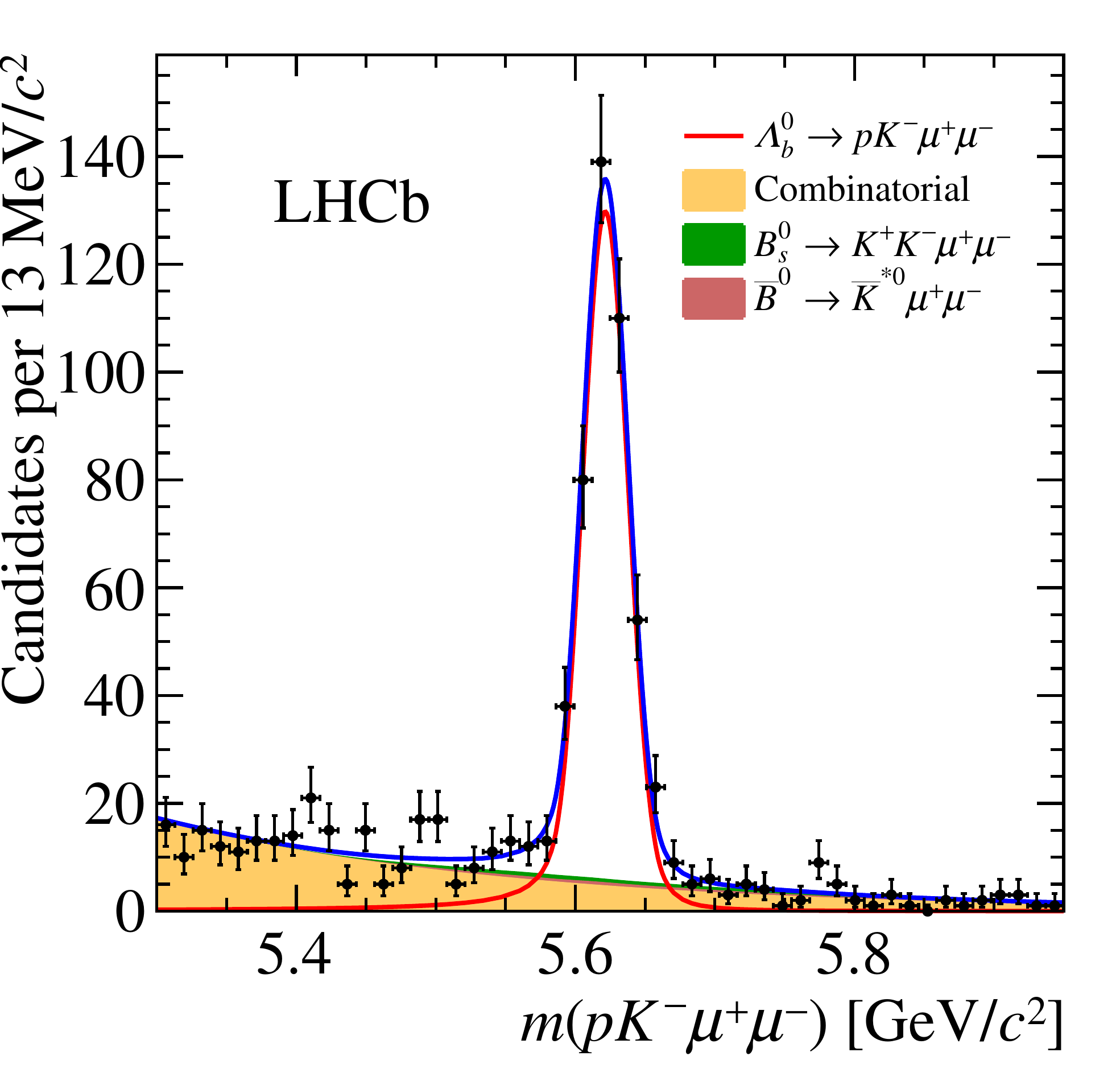}
\includegraphics[width=0.4\textwidth]{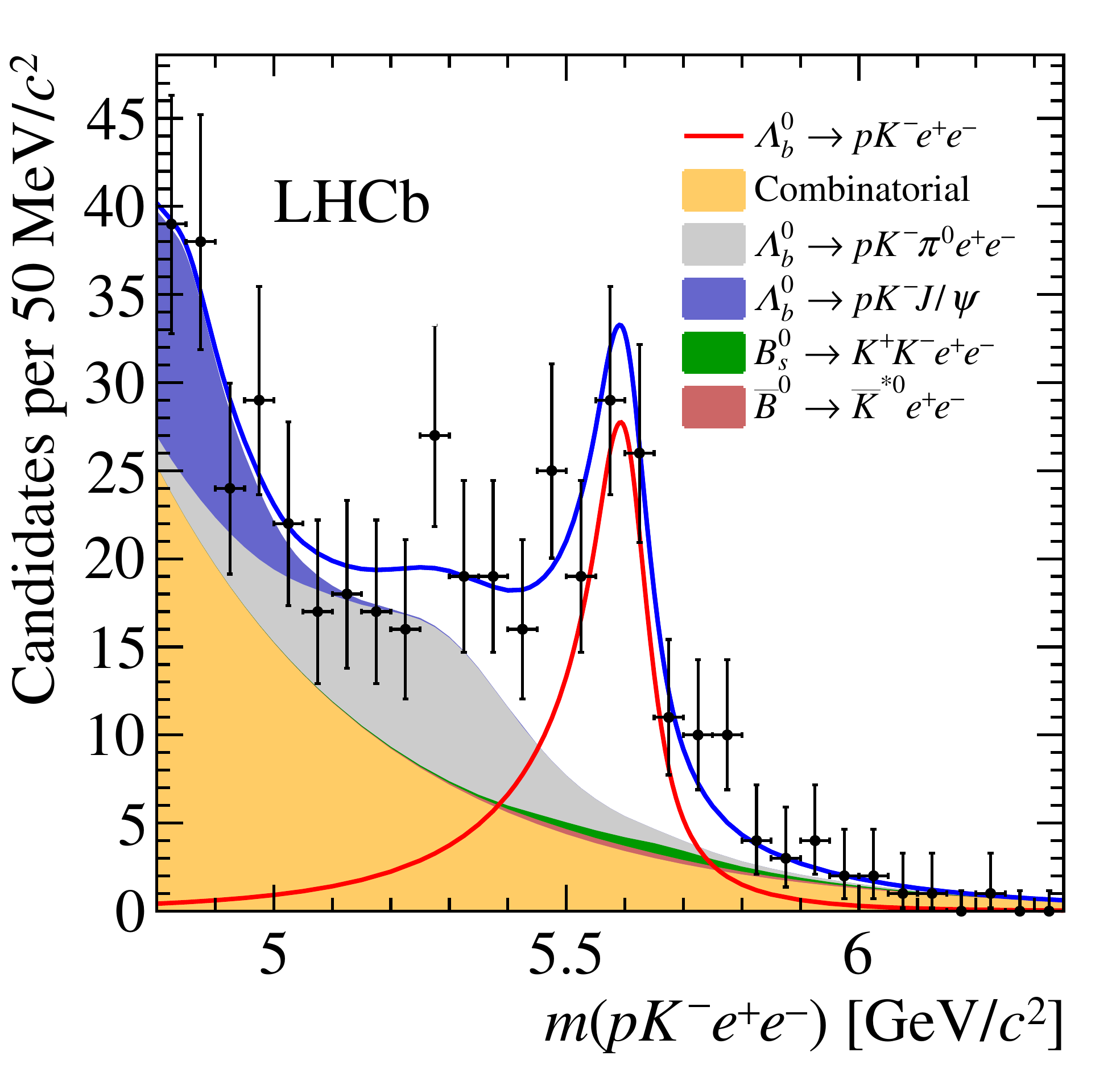}
\end{center}
\vspace{-0.5cm}
\caption{ Invariant mass distributions of (left) $\Lb\to p\Km\mup\mun$ and (right) $\Lb\to p\Km e^+e^-$ candidates. The fit shapes of the $\Lb\to p\Km \ell^+\ell^-$ signals and the main backgrounds are overlaid.}
\label{Fig:LbtopKll_massfits}
\end{figure}

The obtained measurements are $R_K=0.846^{+0.060}_{-0.054}(stat)^{+0.016}_{-0.014}(syst)$ ($1.1<q^2<6\gevgevcccc$); $R_{K^*}=0.66^{+0.11}_{-0.07}(stat)\pm0.03(syst)$ for $0.045<q^2<1.1\gevgevcccc$ and $0.69^{+0.11}_{-0.07}(stat)\pm0.05(syst)$ for $1.1<q^2<6\gevgevcccc$. All these values are systematically below the SM-based predictions by 2.2$\sigma$ to 2.5$\sigma$ \cite{RK_theo}. With the $\Lb\to p\Km\ell^+\ell^-$ decay, a first observation of $\Lb\to p\Km e^+e^-$ is obtained with a similar Run 1 and part of Run 2 data set, as illustrated in Fig.\ref{Fig:LbtopKll_massfits}, leading to the measurement of the ratio $R_{pK}=0.86^{+0.14}_{-0.11}(stat)\pm0.05(syst)$. For all these numbers, the uncertainty will be reduced soon with the addition of the remainder of Run 2 data.

\begin{figure}[htb]%bt\centering
\begin{center}
\includegraphics[width=0.4\textwidth]{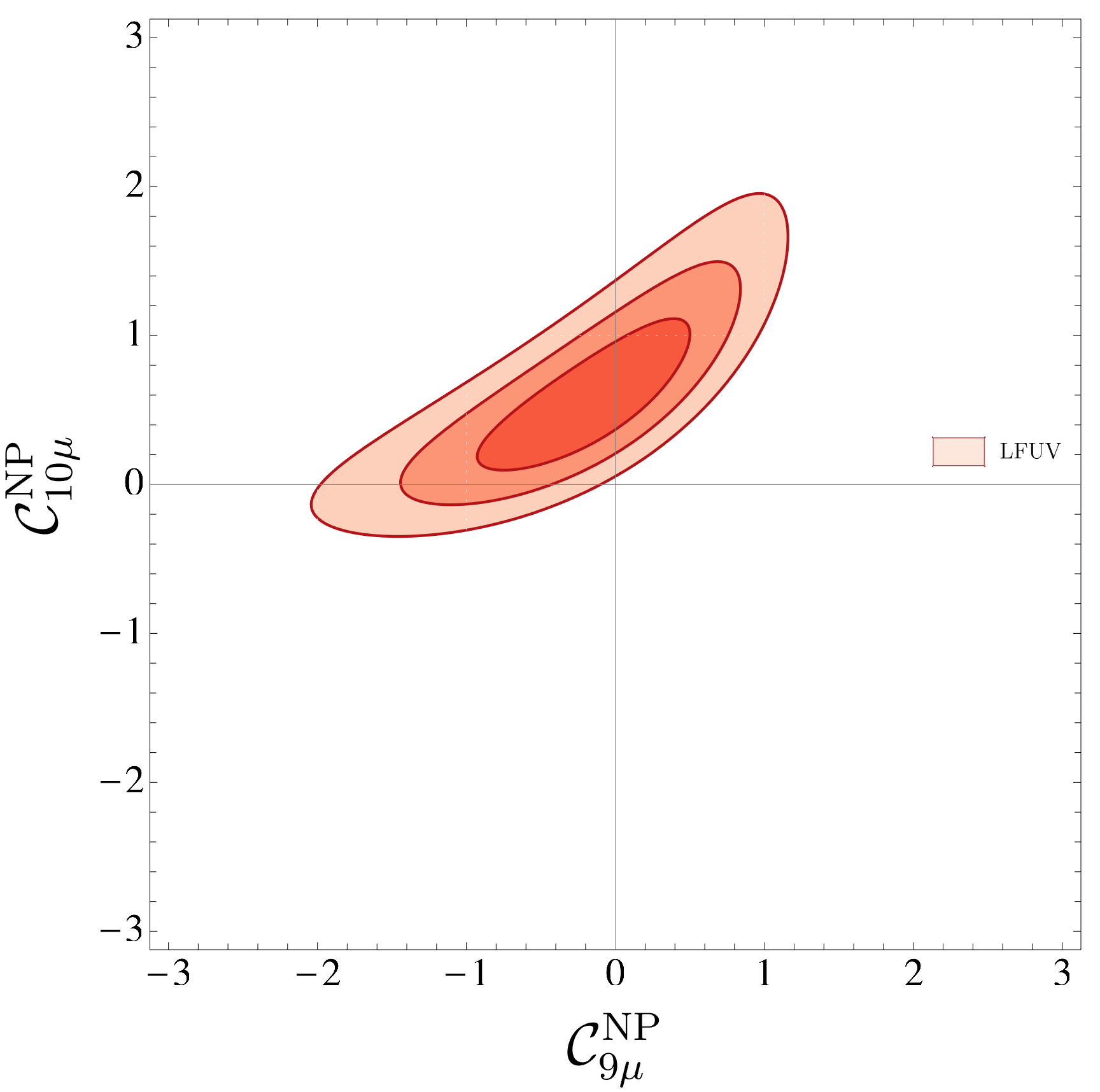}
\includegraphics[width=0.4\textwidth]{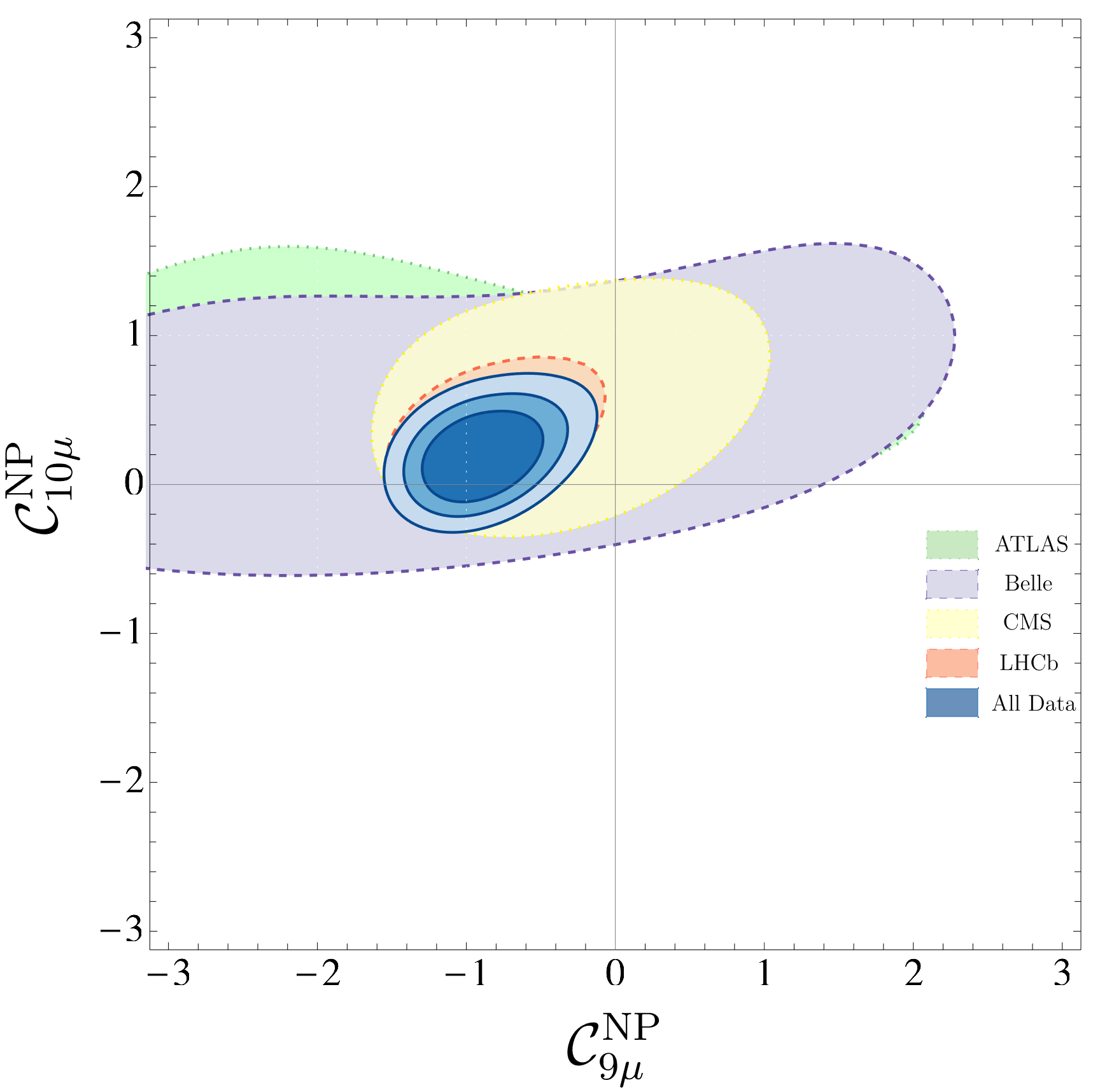}
\end{center}
\vspace{-0.5cm}
\caption{ The 1,2,3 $\sigma$ contours for the New Physics contributions to the $C_9$ and $C_{10}$ Wilson coefficients using (left) only the LFUV data from Belle and LHCb and (right) combining all data from angular analyses and LFUV. The fits are provided in Ref.\cite{C9C10_fit}.}
\label{Fig:C9C10_NP}
\end{figure}

Figure \ref{Fig:C9C10_NP} shows the impact of the $R_{K^{*}}$ measurements by BaBar and LHCb, as well as the combination of the LFUV and angular parameters from all the experiments, on the New Physics contributions to the Wilson coefficients $C_9$ and $C_{10}$, as derived in Ref.\cite{C9C10_fit}. Combining all anomalous data, $C_9$ departs by more than 6$\sigma$ from its SM-based prediction.

\section{$B\to\ell\ell$}

For these purely leptonic modes, the combination of the most recent results of the ATLAS \cite{atlas_bmumu}, CMS \cite{cms_bmumu} and LHCb \cite{lhcb_bmumu} experiments, giving $\mathcal{B}(\Bs\to\mup\mun)= (2.69^{+0.37}_{-0.35})\times10^{-9}$ and $\mathcal{B}(B^0\to\mup\mun)<1.9\times10^{-10}$ at 95\% confidence level (CL), shows that $\Bs\to\mup\mun$ is 2$\sigma$ below the SM-based predictions. First attempts by LHCb of measuring the ditauon \cite{lhcb_btautau} and dielectron \cite{lhcb_bee} modes lead to the results $\mathcal{B}(\Bs\to \tau^+\tau^-)<6.3\times10^{-3}$, $\mathcal{B}(B^0\to \tau^+\tau^-)<2.1\times10^{-3}$, $\mathcal{B}(\Bs\to e^+e^-)<11.2\times10^{-9}$ and $\mathcal{B}(B^0\to e^+e^-)<3.0\times10^{-9}$ at 95\% CL.

\section{LFV searches}

The hints of LFUV in $b \to s\ell\ell$ decays have motivated recent LFV searches, seeking to observe decays of the type $b \to s\ell\ell^\prime$ or $B \to \ell\ell^\prime$. A first study of $\Bp\to\Kp\mu^{\pm}e^{\mp}$ \cite{lhcb_kmue} lead to the establishment of the 95\% CL limits: $\mathcal{B}(\Bp\to\Kp\mu^{-}e^{+})< 9.5\times10^{-9}$ and $\mathcal{B}(\Bp\to\Kp\mu^{+}e^{-})< 8.8\times10^{-9}$. Another analysis on $\Bp\to\Kp\mu^{-}\tau^{+}$ \cite{lhcb_kmutau}, characterized by the original use of the decay $B_{s2}^{*0}\to\Bp\Kp$ to constraint the $\tau$ four-momentum, obtained the less stringent limit $\mathcal{B}(\Bp\to\Kp\mu^{-}\tau^{+})< 4.5\times10^{-5}$. For what concerns the LFV leptonic modes, the decays $B^0_{(s)}\to e^{\pm}\mu^{\mp}$ \cite{lhcb_emu} and $B^0_{(s)}\to \tau^{\pm}\mu^{\mp}$ \cite{lhcb_taumu} have been studied , setting the 95\% CL limits to $\mathcal{B}(B^0_{s}\to e^{\pm}\mu^{\mp})< 6.3\times10^{-9}$, $\mathcal{B}(B^0\to e^{\pm}\mu^{\mp})< 1.3\times10^{-9}$, $\mathcal{B}(B^0_{s}\to \tau^{\pm}\mu^{\mp})< 4.2\times10^{-5}$ and $\mathcal{B}(B^0\to \tau^{\pm}\mu^{\mp})< 1.4\times10^{-5}$.

%On the LFV front, LHCb published results on the search for the decays $B^0_{(s)}\to \tau^{\pm}\mu^{\mp}$ \cite{lhcb_taumu}, with the 95\% CL limits being $\mathcal{B}(B^0_{s}\to \tau^{\pm}\mu^{\mp})< 4.2\times10^{-5}$ and $\mathcal{B}(B^0\to \tau^{\pm}\mu^{\mp})< 1.4\times10^{-5}$.

\section{Prospects and summary}

Most analyses presented have been published on a partial LHCb data set and are currently being updated. The second column of Table \ref{Tab:RX_prec} shows the expected precisions for the $R_X$ measurements for the full Run1+Run2 statistics. 

\begin{table}
\centering
	\caption{Expected precisions for $R_X = \mathcal{B}(H_b\to X \mu^+\mu^-)/\mathcal{B}(H_b\to X e^+e^-)$. The numbers are taken from Ref.\cite{RX_UpgradeII}}\label{Tab:RX_prec}
	\begin{tabular}{l|cccc}
        \hline
	$R_X$                     & Run 1\&2 (9\invfb) & Run 3 (23\invfb)& Run 4 (50\invfb)& Run 5 (300\invfb)\\
	\hline
        $R_K$ &  0.043 &  0.025 &  0.017 &  0.007\\
        $R_{K^{*0}}$ &  0.052 &  0.031 &  0.020 &  0.008\\
        $R_{\phi}$ &  0.130 &  0.076 &  0.050 &  0.020\\
        $R_{pK}$ &  0.105 &  0.061 &  0.041 &  0.016\\
        $R_{\pi}$ &  0.302 &  0.176 &  0.117 &  0.047\\
        \hline\hline
	\end{tabular}
\end{table}

\begin{figure}[htb]%bt\centering
\begin{center}
\includegraphics[width=0.6\textwidth]{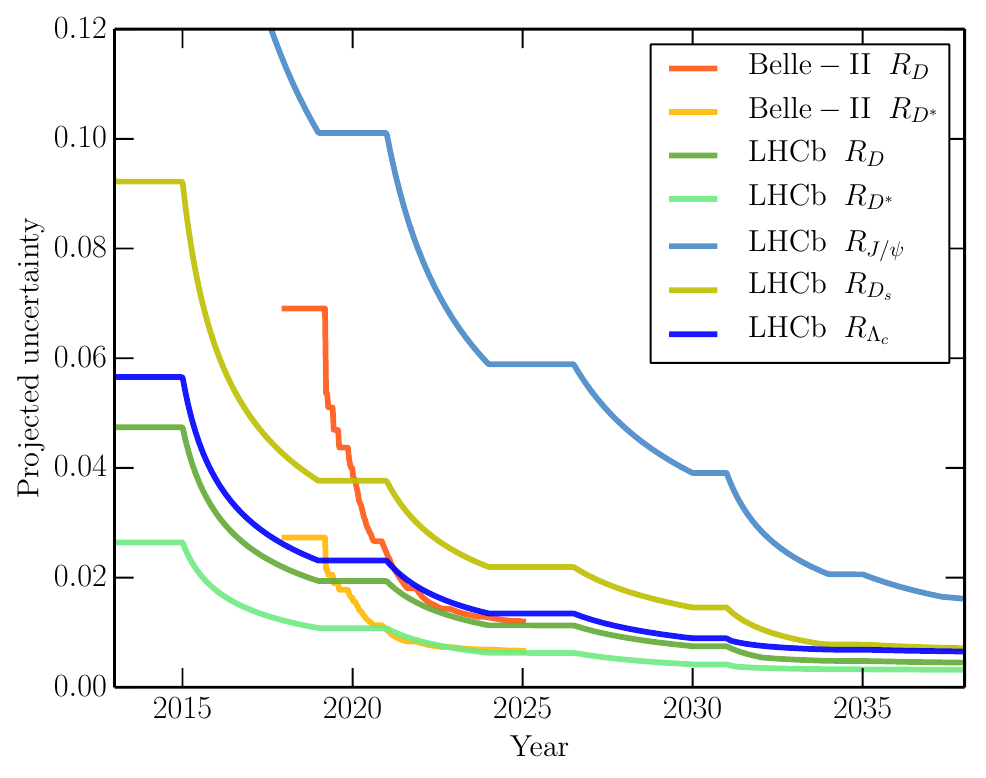}
\end{center}
\caption{Evolution of the sensitivity on the semileptonic ratios $R(H_c)$ as reported in Ref.\cite{RD_UpgradeII}.}
\label{Fig:RD_UpgradeII}
\end{figure}

On the longer term, the third, fourth and fifth columns of Table \ref{Tab:RX_prec} show the evolution of the expected sensitivities for the future runs of data taking. For the tree semileptonic decays, the LFU ratios $R(D^0)$, $R(D^+)$, $R(D_s^{(*)})$,  $R(\Lambda_c^{(*)})$, $R(\jpsi)$ and $R(p)$ (from $\Lb\to p\tau\nu$) are foreseen during the first phase of Run 3. Figure \ref{Fig:RD_UpgradeII} shows the evolution of the $R(H_c)$ ratios throughout the periods of data taking.

The available results on the anomalies in the $b$-hadron decays show a combination of deviations, which has triggered an intense activity on the phenomenological side in studies aiming at constraining the Wilson coefficients and probing possible contributions of New Physics \cite{C9C10_fit,flavio,straub}. The 6 to 7$\sigma$ deviation from SM derived for $C_9$ is subject to interpretations, which try to account for what is observed in both $b\to c$ tree transitions and $b\to s$ loop decays. Explanations based on vector Leptoquarks \cite{vector_leptoquarks} and the ``4321'' model \cite{4321} have become popular. Any explored paradigm will have to satisfy the constraints of the \Bs meson mixing and \Bc meson lifetime.

\end{document}

%% file: preamble.tex
\usepackage{microtype}
\usepackage{lineno}  % for line numbering during review
\usepackage{xspace} % To avoid problems with missing or double spaces after
\usepackage{caption} %these three command get the figure and table captions automatically small

\usepackage{graphicx,subfigure}  % to include figures (can also use other packages)
\usepackage{color}
\usepackage{colortbl}
\graphicspath{{./figs/}} % Make Latex search fig subdir for figures

\usepackage{amsmath} % Adds a large collection of math symbols
\usepackage{amssymb}
\usepackage{amsfonts}
\usepackage{upgreek} % Adds in support for greek letters in roman typeset

\newcommand*\patchAmsMathEnvironmentForLineno[1]{%
\expandafter\let\csname old#1\expandafter\endcsname\csname #1\endcsname
\expandafter\let\csname oldend#1\expandafter\endcsname\csname
end#1\endcsname
 \renewenvironment{#1}%
   {\linenomath\csname old#1\endcsname}%
   {\csname oldend#1\endcsname\endlinenomath}%
}
\newcommand*\patchBothAmsMathEnvironmentsForLineno[1]{%
  \patchAmsMathEnvironmentForLineno{#1}%
  \patchAmsMathEnvironmentForLineno{#1*}%
}
\AtBeginDocument{%
\patchBothAmsMathEnvironmentsForLineno{equation}%
\patchBothAmsMathEnvironmentsForLineno{align}%
\patchBothAmsMathEnvironmentsForLineno{flalign}%
\patchBothAmsMathEnvironmentsForLineno{alignat}%
\patchBothAmsMathEnvironmentsForLineno{gather}%
\patchBothAmsMathEnvironmentsForLineno{multline}%
\patchBothAmsMathEnvironmentsForLineno{eqnarray}%
}

\usepackage{url}    % Hyperlinks in references
\input{lhcb-symbols-def} % Add in the predefined LHCb symbols

\usepackage{cite} % Allows for ranges in citations
\usepackage{mciteplus}

%% file: lhcb-symbols-def.tex
\usepackage{xspace} 
\usepackage{upgreek}

\def\MagUp {\mbox{\em Mag\kern -0.05em Up}\xspace}

\ifthenelse{\boolean{uprightparticles}}%
{

 \def\Pmu         {\ensuremath{\upmu}\xspace}

 \def\Ppi         {\ensuremath{\uppi}\xspace}

 \def\Ppsi        {\ensuremath{\uppsi}\xspace}

 \def\PDelta      {\ensuremath{\Delta}\xspace}                 
 \def\PXi      {\ensuremath{\Xi}\xspace}                 
 \def\PLambda      {\ensuremath{\Lambda}\xspace}                 
 \def\PSigma      {\ensuremath{\Sigma}\xspace}                 
 \def\POmega      {\ensuremath{\Omega}\xspace}                 
 \def\PUpsilon      {\ensuremath{\Upsilon}\xspace}                 
 
 \def\PB      {\ensuremath{\mathrm{B}}\xspace}                 
                  
 \def\PD      {\ensuremath{\mathrm{D}}\xspace}

 \def\PJ      {\ensuremath{\mathrm{J}}\xspace}                 
 \def\PK      {\ensuremath{\mathrm{K}}\xspace}

 \def\Pb      {\ensuremath{\mathrm{b}}\xspace}                 
 \def\Pc      {\ensuremath{\mathrm{c}}\xspace}

 \def\Pi      {\ensuremath{\mathrm{i}}\xspace}

 \def\Ps      {\ensuremath{\mathrm{s}}\xspace}

}
{

 \def\Pmu         {\ensuremath{\mu}\xspace}

 \def\Ppi         {\ensuremath{\pi}\xspace}

 \def\Ppsi        {\ensuremath{\psi}\xspace}                 
                  
 \mathchardef\PDelta="7101
 \mathchardef\PXi="7104
 \mathchardef\PLambda="7103
 \mathchardef\PSigma="7106
 \mathchardef\POmega="710A
 \mathchardef\PUpsilon="7107
                  
 \def\PB      {\ensuremath{B}\xspace}                 
                  
 \def\PD      {\ensuremath{D}\xspace}

 \def\PJ      {\ensuremath{J}\xspace}                 
 \def\PK      {\ensuremath{K}\xspace}

 \def\Pb      {\ensuremath{b}\xspace}                 
 \def\Pc      {\ensuremath{c}\xspace}

 \def\Pi      {\ensuremath{i}\xspace}

 \def\Ps      {\ensuremath{s}\xspace}

}

\makeatletter
\ifcase \@ptsize \relax% 10pt
  \newcommand{\miniscule}{\@setfontsize\miniscule{4}{5}}% \tiny: 5/6
\or% 11pt
  \newcommand{\miniscule}{\@setfontsize\miniscule{5}{6}}% \tiny: 6/7
\or% 12pt
  \newcommand{\miniscule}{\@setfontsize\miniscule{5}{6}}% \tiny: 6/7
\fi
\makeatother

\DeclareRobustCommand{\optbar}[1]{\shortstack{{\miniscule (\rule[.5ex]{1.25em}{.18mm})}
  \\ [-.7ex] $#1$}}

\def\mup        {{\ensuremath{\Pmu^+}}\xspace}
\def\mun        {{\ensuremath{\Pmu^-}}\xspace} % muon negative (\mum is taken)

\def\squark    {{\ensuremath{\Ps}}\xspace}

\def\cquark    {{\ensuremath{\Pc}}\xspace}

\def\bquark    {{\ensuremath{\Pb}}\xspace}

\def\pion   {{\ensuremath{\Ppi}}\xspace}

\def\pip    {{\ensuremath{\pion^+}}\xspace}
\def\pim    {{\ensuremath{\pion^-}}\xspace}

\def\kaon    {{\ensuremath{\PK}}\xspace}
  \def\Kbar    {{\kern 0.2em\overline{\kern -0.2em \PK}{}}\xspace}

\def\KorKbar    {\kern 0.18em\optbar{\kern -0.18em K}{}\xspace}

\def\Kp      {{\ensuremath{\kaon^+}}\xspace}
\def\Km      {{\ensuremath{\kaon^-}}\xspace}

  \def\Dbar    {{\kern 0.2em\overline{\kern -0.2em \PD}{}}\xspace}

\def\DorDbar    {\kern 0.18em\optbar{\kern -0.18em D}{}\xspace}

\def\B       {{\ensuremath{\PB}}\xspace}
\def\Bbar    {{\ensuremath{\kern 0.18em\overline{\kern -0.18em \PB}{}}}\xspace}

\def\BorBbar    {\kern 0.18em\optbar{\kern -0.18em B}{}\xspace}

\def\Bu      {{\ensuremath{\B^+}}\xspace}

\def\Bp      {{\ensuremath{\Bu}}\xspace}

\def\Bs      {{\ensuremath{\B^0_\squark}}\xspace}

\def\Bc      {{\ensuremath{\B_\cquark^+}}\xspace}

\def\jpsi     {{\ensuremath{{\PJ\mskip -3mu/\mskip -2mu\Ppsi\mskip 2mu}}}\xspace}
\def\psitwos  {{\ensuremath{\Ppsi{(2S)}}}\xspace}

  \def\Y#1S{\ensuremath{\PUpsilon{(#1S)}}\xspace}% no space before {...}!

\def\Lz          {{\ensuremath{\PLambda}}\xspace}
\def\Lbar        {{\ensuremath{\kern 0.1em\overline{\kern -0.1em\PLambda}}}\xspace}
\def\LorLbar    {\kern 0.18em\optbar{\kern -0.18em \PLambda}{}\xspace}

\def\Lb      {{\ensuremath{\Lz^0_\bquark}}\xspace}

\def\to                 {\ensuremath{\rightarrow}\xspace}

\def\AT#1     {\ensuremath{A_{\mathrm{T}}^{#1}}\xspace}           % 2

\def\C#1      {\ensuremath{\mathcal{C}_{#1}}\xspace}                       % 9
\def\Cp#1     {\ensuremath{\mathcal{C}_{#1}^{'}}\xspace}                    % 7
\def\Ceff#1   {\ensuremath{\mathcal{C}_{#1}^{\mathrm{(eff)}}}\xspace}        % 9  
\def\Cpeff#1  {\ensuremath{\mathcal{C}_{#1}^{'\mathrm{(eff)}}}\xspace}       % 7
\def\Ope#1    {\ensuremath{\mathcal{O}_{#1}}\xspace}                       % 2
\def\Opep#1   {\ensuremath{\mathcal{O}_{#1}^{'}}\xspace}                    % 7

\newcommand{\tev}{\ifthenelse{\boolean{inbibliography}}{\ensuremath{~T\kern -0.05em eV}}{\ensuremath{\mathrm{\,Te\kern -0.1em V}}}\xspace}
\newcommand{\gev}{\ensuremath{\mathrm{\,Ge\kern -0.1em V}}\xspace}
\newcommand{\mev}{\ensuremath{\mathrm{\,Me\kern -0.1em V}}\xspace}
\newcommand{\kev}{\ensuremath{\mathrm{\,ke\kern -0.1em V}}\xspace}
\newcommand{\ev}{\ensuremath{\mathrm{\,e\kern -0.1em V}}\xspace}
\newcommand{\gevc}{\ensuremath{{\mathrm{\,Ge\kern -0.1em V\!/}c}}\xspace}
\newcommand{\mevc}{\ensuremath{{\mathrm{\,Me\kern -0.1em V\!/}c}}\xspace}
\newcommand{\gevcc}{\ensuremath{{\mathrm{\,Ge\kern -0.1em V\!/}c^2}}\xspace}
\newcommand{\gevgevcccc}{\ensuremath{{\mathrm{\,Ge\kern -0.1em V^2\!/}c^4}}\xspace}
\newcommand{\mevcc}{\ensuremath{{\mathrm{\,Me\kern -0.1em V\!/}c^2}}\xspace}

\def\invfb   {\ensuremath{\mbox{\,fb}^{-1}}\xspace}

\def\gsim{{~\raise.15em\hbox{$>$}\kern-.85em
          \lower.35em\hbox{$\sim$}~}\xspace}
\def\lsim{{~\raise.15em\hbox{$<$}\kern-.85em
          \lower.35em\hbox{$\sim$}~}\xspace}

\def\tell1  {TELL1\xspace}
\def\ukl1   {UKL1\xspace}

%% file: main.bbl
\begin{thebibliography}{99}
\bibitem{Buchalla} G.~Buchalla {\em et~al.}, Rev.\ Mod.\ Phys.\ \textbf{68} (1996) 1125-1144
\bibitem{Higgsrev_2HDM} J.F. Gunion, E.H. Haber, G.L. Kane, S. Dawson, {\it The Higgs Hunter's Guide}, Front.Phys.\textbf{80} (2000) 1-404
\bibitem{Dstar_MuMode} LHCb collaboration, R.~Aaij {\em et~al.}, Phys.\ Rev.\ Lett. \textbf{115} (2015) 111803
\bibitem{Dstar_HadMode} LHCb collaboration, R.~Aaij {\em et~al.}, Phys.\ Rev.\ Lett. \textbf{120} (2018) 171802
\bibitem{BDT} L.~Breiman, J.H.~Friedman, R.A.~Olshen and C.J.~Stone, {\it Classification and regression trees}, Wadsworth international group (1984) Belmont, California, USA
\bibitem{hflav} Heavy Flavor Averaging Group, Y. Amhis {\em et~al.}, arXiv:1909.12524, latest results and plots available at \href{https://hflav.web.cern.ch/}{{\texttt{https://hflav.web.cern.ch/}}}
\bibitem{belle_rd_2019} Belle Collaboration, A.~Abdesselam {\em et~al.}, arXiv:1904.08794
\bibitem{rd_theory} D. Bigi, P. Gambino, Phys.\ Rev.\ \textbf{D94} (2016) 094008; F.Bernlochner {\em et~al.}, Phys.\ Rev.\ \textbf{D95} (2017) 115008 ; D.Bigi {\em et~al.}, JHEP \textbf{1711} (2017) 061; S.Jaiswal {\em et~al.}, JHEP \textbf{1712} (2017) 060
\bibitem{RJpsi} LHCb collaboration, R.~Aaij {\em et~al.}, Phys.\ Rev.\ Lett. \textbf{120} (2018) 121801
\bibitem{RJpsi_theory} A.Yu.Anisimov {\em et~al.}, Phys.\ Lett. \textbf{B452} (1999) 129; M.A.Ivanov {\em et~al.}, Phys.\ Rev.\ \textbf{D73} (2006) 054024; E.Hernandez {\em et~al.}, Phys.\ Rev.\ \textbf{D74} (2006) 074008
\bibitem{lhcb_xmumu} LHCb collaboration, R.~Aaij {\em et~al.}, a: JHEP \textbf{06} (2014) 133; b: JHEP \textbf{09} (2015) 179; c: PRL 125 (2020) 011802; d: JHEP \textbf{06} (2015) 115
\bibitem{lhcb_lzmumu_ang} LHCb collaboration, R.~Aaij {\em et~al.}, JHEP \textbf{09} (2018) 146
\bibitem{descotes} Descotes-Genon {\em et~al.}, JHEP \textbf{05} (2013) 137
\bibitem{lhcb_rkplus_ratio}  LHCb collaboration, R.~Aaij {\em et~al.}, Phys.\ Rev.\ Lett. \textbf{122} (2019) 191801
\bibitem{lhcb_rkstar_ratio}  LHCb collaboration, R.~Aaij {\em et~al.}, JHEP \textbf{08} (2017) 055
\bibitem{lhcb_rpk_ratio}  LHCb collaboration, R.~Aaij {\em et~al.}, JHEP \textbf{05} (2020) 040
\bibitem{RK_theo} Non-exhaustive: C. Bobeth {\em et~al.} JHEP \textbf{07} (2007) 040; M.Bordone {\em et~al.}, Eur.\ Phys.\ J.\ \textbf{C76} (2016) 440; W. Altmannshofer {\em et~al.}, Phys.\ Rev.\ \textbf{D96} (2017) 055008
\bibitem{C9C10_fit} M. Alguero {\em et~al.}, Eur. Phys. J. \textbf{C79} (2019) 8, 714; Eur. Phys. J.\textbf{C80} (2020) 6, 511 (addendum).
\bibitem{atlas_bmumu} ATLAS collaboration, ATLAS-CONF-2020-049
\bibitem{cms_bmumu} CMS collaboration, CMS-PAS-BPH-20-003
\bibitem{lhcb_bmumu} LHCb collaboration, LHCb-CONF-2020-002
\bibitem{lhcb_btautau} LHCb collaboration,  R.~Aaij {\em et~al.}, Phys.\ Rev.\ Lett. \textbf{118} (2017) 251802
\bibitem{lhcb_bee} LHCb collaboration,  R.~Aaij {\em et~al.}, Phys.\ Rev.\ Lett. \textbf{124} (2020) 211802
\bibitem{lhcb_kmue} LHCb collaboration, R.~Aaij {\em et~al.}, Phys.\ Rev.\ Lett. \textbf{123} (2019) 241802
\bibitem{lhcb_kmutau} LHCb collaboration, R.~Aaij {\em et~al.}, JHEP \textbf{06} (2020) 129
\bibitem{lhcb_emu}  LHCb collaboration, R.~Aaij {\em et~al.}, JHEP \textbf{03} (2018) 078
\bibitem{lhcb_taumu} LHCb collaboration, R.~Aaij {\em et~al.}, Phys.\ Rev.\ Lett. \textbf{123} (2019) 211801
\bibitem{RX_UpgradeII} LHCb collaboration, R.~Aaij {\em et~al.}, arXiv:1808.08865, LHCB-PUB-2018-009, CERN-LHCC-2018-027
\bibitem{RD_UpgradeII} S.~Bifani {\em et~al.}, J. Phys. G: Nucl. Part. Phys. 46 (2019) 023001
\bibitem{straub} J. Aebischer {\em et~al.}, arXiv:1903.10434
\bibitem{flavio} D.~M.~Straub, {\tt flavio} package, arXiv:1810.08132
\bibitem{vector_leptoquarks} D. Buttazzo {\em et~al.}, JHEP 11 (2017) 044
\bibitem{4321} L. Di Luzio {\em et~al.}, Phys. Rev. \textbf{D96}, (2017) 115011; JHEP \textbf{11} (2018) 081


\end{thebibliography}
